\documentclass[preprint,prd,nofootinbib,preprintnumbers]{revtex4-1}
\usepackage{amssymb,amsmath,bm, mathrsfs}
\usepackage{amsfonts}
\usepackage{amstext}
\usepackage{array, multirow, graphicx, float}
\usepackage{color}
\usepackage[normalem]{ulem}
\usepackage{slashed}
\usepackage{hyperref}
\usepackage{textcomp}

\begin{document}

\title{Predictive Dirac and Majorana Neutrino Mass Textures from $SU(6)$ Grand Unified Theories}

\author{Zackaria Chacko$^a$, P. S. Bhupal Dev$^b$, Rabindra N. Mohapatra$^a$, Anil Thapa$^c$}
\affiliation{$^a$Maryland Center for Fundamental Physics, Department of
Physics, University of Maryland, College Park, MD 20740, USA}
\affiliation{$^b$ Department of Physics and McDonnell Center for the Space Sciences, Washington University, St. Louis, MO 63130, USA}
\affiliation{$^c$ Department of Physics, Oklahoma State University, Stillwater, OK 74078, USA}

 \begin{abstract}
 We present simple and predictive realizations of neutrino masses in 
theories based on the $SU(6)$ grand unifying group. At the level of the 
lowest-dimension operators, this class of models predicts a 
skew-symmetric flavor structure for the Dirac mass term of the 
neutrinos. In the case that neutrinos are Dirac particles, the 
lowest-order prediction of this construction is then one massless 
neutrino and two degenerate massive neutrinos. Higher-dimensional 
operators suppressed by the Planck scale perturb this spectrum, allowing 
a good fit to the observed neutrino mass matrix. A firm prediction of 
this construction is an inverted neutrino mass spectrum with the 
lightest neutrino hierarchically lighter than the other two, so that the 
sum of neutrino masses lies close to the lower bound for an inverted 
hierarchy. In the alternate case that neutrinos are Majorana particles, 
the mass spectrum can be either normal or inverted. However, the 
lightest neutrino is once again hierarchically lighter than the other 
two, so that the sum of neutrino masses is predicted to lie close to the 
corresponding lower bound for the normal or inverted hierarchy. Near 
future cosmological measurements will be able to test the predictions of 
this scenario for the sum of neutrino masses. In the case of Majorana 
neutrinos that exhibit an inverted hierarchy, future neutrinoless double 
beta experiments can provide a complementary probe.
 \end{abstract}

\maketitle

\section{Introduction}

 Multiple neutrino oscillation experiments over the past two decades 
have conclusively established that neutrinos have non-vanishing 
masses~\cite{Tanabashi:2018oca}, thereby providing concrete evidence of 
new physics beyond the Standard Model (SM). However, although these 
experiments have measured the neutrino mass splittings and mixing 
angles, the actual values of the neutrino masses still remain unknown. 
In particular, it is not known whether the neutrino mass spectrum 
exhibits a normal or inverted hierarchy. Several medium and 
long-baseline neutrino oscillation experiments have been proposed to 
settle this issue~\cite{Qian:2015waa}. At present, the important 
question of whether neutrinos are Dirac or Majorana fermions also 
remains unanswered. Future neutrinoless double beta decay 
($0\nu\beta\beta$) experiments may be able to resolve this 
question~\cite{Dolinski:2019nrj}.

 Grand unification~\cite{Pati:1974yy,Georgi:1974sy,Langacker:1980js} is 
one of the most attractive proposals for physics beyond the SM. In these 
theories, the strong, weak and electromagnetic interactions of the SM 
are unified into a larger grand unifying group. The fermions of the SM 
are embedded into representations of this bigger group, with the result 
that quarks and leptons are also unified into the same multiplets. These 
representations often contain additional SM singlets, which can 
naturally serve the role of right-handed neutrinos in the generation of 
neutrino masses. The fact that the SM quarks and leptons are now 
embedded together in the same multiplets often leads to relations 
between the masses of the different SM fermions~\cite{Buras:1977yy}. If 
these multiplets also contain right-handed neutrinos, these theories can 
impose restrictions on the form of the neutrino mass matrix, leading to 
predictions for the neutrino masses. Familiar examples of unified theories 
that can relate the masses of the neutrinos to those of the charged 
fermion include the Pati-Salam~\cite{Pati:1974yy} and 
$SO(10)$~\cite{Georgi:1974my, Fritzsch:1974nn} gauge groups.

 In this paper we explore a class of models based on the $SU(6)$ grand 
unified theory (GUT)~\cite{Kim:1981jw,Fukugita:1981gn} that lead to 
sharp predictions for the neutrino mass spectrum. In these theories, the 
right-handed neutrino emerges from the same multiplet as the lepton 
doublet of the SM. A natural consequence of this construction is that, 
at the level of the lowest-dimension terms, the Dirac mass term for the 
neutrinos is skew-symmetric in flavor space, so that the determinant of 
the Dirac mass matrix vanishes. If neutrinos are Dirac particles that 
obtain their masses from this term, then, in the absence of corrections 
to this form from terms of higher dimension, the neutrino mass spectrum 
consists of two degenerate species and a massless one. Once 
higher-dimensional terms suppressed by the Planck scale $M_{\rm Pl}$ are 
included, this class of models can easily reproduce the observed 
spectrum of neutrino masses and mixings. A firm prediction of this 
construction is that the spectrum of neutrino masses is inverted, with 
the lightest neutrino hierarchically lighter than the other two. Then 
the sum of neutrino masses is predicted to lie close to the lower bound 
of 0.10 eV set by the observed mass splittings in the case of an 
inverted hierarchy. Future precision cosmological experiments, such as 
LSST~\cite{Abell:2009aa}, Euclid~\cite{Amendola:2016saw}, 
DESI~\cite{Aghamousa:2016zmz}, the Simons 
Observatory~\cite{Ade:2018sbj}, and CMB-S4~\cite{Abazajian:2019eic}, 
that have the required sensitivity to the sum of neutrino masses will be 
able to test this striking prediction. The final phase of 
Project-8~\cite{Esfahani:2017dmu}, with an expected sensitivity of 0.04 
eV to the absolute electron neutrino mass, will also be able to test 
this scenario. Similarly, future large-scale long-baseline neutrino 
oscillation experiments, such as Hyper-K~\cite{Abe:2018uyc} and 
DUNE~\cite{Abi:2018dnh}, will be able to test the prediction regarding 
the inverted nature of the mass spectrum.

It is well-established that there is a lower bound on the light neutrino 
contribution to the $0\nu\beta\beta$ process in the case of Majorana 
neutrinos that exhibit an inverted mass-hierarchy~\cite{Vissani:1999tu, 
Bilenky:2001rz}. In particular, it has been pointed out that if 
long-baseline neutrino experiments determine that the neutrino mass 
hierarchy is inverted, while no signal is observed in $0\nu\beta\beta$ 
down to the effective Majorana neutrino mass $ m_{ee}\lesssim 0.03$ eV, 
then this would constitute compelling evidence that neutrinos are Dirac 
rather than Majorana fermions~\cite{Mohapatra:2005wg}. The model we 
present here is an example of a GUT framework that can naturally 
accommodate such a scenario.

If, in addition to the skew-symmetric Dirac mass term, there is also a 
large Majorana mass term for the right-handed neutrinos, the neutrinos 
will be Majorana particles. In this scenario, the skew-symmetric nature 
of the Dirac mass term implies that the lightest neutrino is massless, 
up to small corrections from higher-dimensional operators. 
In contrast to the case of Dirac neutrinos discussed above, the spectrum 
of neutrino masses can now exhibit either a normal or inverted 
hierarchy. However, the lightest neutrino is still predicted to be 
hierarchically lighter than the other two, so that for both normal and 
inverted hierarchies the sum of neutrino masses is predicted to lie 
close to the corresponding lower bound dictated by the observed mass 
splittings, i.e. 0.06 eV for the normal case and 0.10 eV for the 
inverted. This is a prediction that can be tested by future cosmological 
observations once long-baseline experiments have determined whether the 
spectrum is normal or inverted. In addition, these predictions for the 
sum of neutrino masses translate into upper and lower bounds on the 
$0\nu\beta\beta$ rate for each of the normal and inverted cases, with 
important implications for future $0\nu\beta\beta$ experiments. In our 
analysis, we explore both the Dirac and Majorana possibilities in detail 
and obtain realistic fits to the observed masses and mixings.

 To understand the origin of the prediction that the Dirac mass term for 
the neutrinos is skew-symmetric, we first consider the minimal grand 
unifying symmetry, namely $SU(5)$~\cite{Georgi:1974sy}. In this class 
of theories the $SU(5)$ grand unifying symmetry is broken at the 
unification scale, $M_{\rm GUT} \sim 10^{16}$ GeV, down to the SM gauge 
groups. In simple models based on $SU(5)$, all the SM fermions in a 
single generation arise from the ${\bf \bar{5}}$ and ${\bf 10}$ 
representations. The ${\bf \bar{5}}$ is the anti-fundamental 
representation while the ${\bf 10}$ is the tensor representation with 
two antisymmetric indices. The Higgs field of the SM is contained in the 
fundamental representation, the ${\bf 5}$. The up-type quark masses 
arise from Yukawa couplings of the schematic form $\epsilon^{\kappa 
\lambda \mu \nu \rho} {\bf 5_{\rm H}}_\kappa {\bf 10}_{\lambda \mu} {\bf 
10}_{\nu \rho}$, where ${\bf 5_{\rm H}}$ contains the SM Higgs, 
$\epsilon^{\kappa \lambda \mu \nu \rho}$ is the 5-dimensional 
antisymmetric Levi-Civita tensor, and the Greek letters represent 
$SU(5)$ indices. Similarly, the down-type quark and charged lepton 
masses arise from Yukawa couplings of the form ${\bf 5^{\dagger}_{\rm 
H}}^{\mu} {\bf 10}_{\mu \nu} {\bf \bar{5}}^{\nu}$. Although attractive 
and elegant, the minimal $SU(5)$ model does not contain SM singlets that 
can play the role of right-handed neutrinos, and does not make 
predictions regarding the neutrino masses. Simple extensions of minimal 
$SU(5)$ to $SU(6)$, however, do contain natural candidates for the role 
of right-handed neutrinos and also allow for elegant solutions to the 
doublet-triplet splitting problem~\cite{Sen:1984aq,Barr:1997pt, 
Berezhiani:1989bd,Barbieri:1993wz,Dvali:1993yf,Chacko:1998zz}.

In the simplest extension of $SU(5)$ to $SU(6)$, the SM fermions emerge 
from the ${\bf \bar{6}}$ and ${\bf 15}$ representations. While the ${\bf 
\bar{6}}$ is the antifundamental representation of $SU(6)$, the ${\bf 
15}$ is the tensor representation with two antisymmetric indices. Under 
the $SU(5)$ subgroup of $SU(6)$, these representations decompose as 
${\bf 15} \rightarrow {\bf 10} + {\bf 5}$ and ${\bf \bar{6}} \rightarrow 
{\bf \bar{5}} + {\bf 1}$, and can be seen to contain particles with the 
quantum numbers of the SM fermions. But now, in addition, the singlet of 
$SU(5)$ contained in the ${\bf \bar{6}}$ representation is a natural 
candidate to play the role of the right-handed neutrino. If the SM Higgs 
emerges from the fundamental representation of $SU(6)$, the down-type 
quarks and charged leptons can obtain masses from terms of the schematic 
form ${\bf {6^{\dagger}}_{\rm H}}^{\mu} {\bf 15}_{\mu \nu} {\bf 
{\bar{6}}}^{\nu}$. However, with this set of representations it is not 
possible to obtain masses for the up-type quarks of the SM at the 
renormalizable level. This presents a problem because the top Yukawa 
coupling is large.

One possible solution to this problem, first explored in 
Refs.~\cite{Barbieri:1994kw,Berezhiani:1995dt}, is that the 
third-generation up-type quarks emerge in part from the ${\bf 20}$ of 
$SU(6)$, which is the tensor representation with three antisymmetric 
indices. This decomposes as ${\bf 20} \rightarrow {\bf 10} + {\bf 
\overline{10}}$ under $SU(5)$. This allows the third-generation up-type 
quarks to obtain their masses from a renormalizable term of the form 
$\epsilon^{\kappa \lambda \mu \nu \rho \sigma} {\bf {6_{\rm H}}}_\kappa 
{\bf {15}}_{\lambda \mu} {\bf {20}}_{\nu \rho \sigma}$. 
Nonrenormalizable operators suffice to generate masses for the up-type 
quarks of the lighter two generations.

The problem of the top quark mass in $SU(6)$ GUTs admits an alternative 
solution if electroweak symmetry is broken by two light Higgs doublets 
rather than one, so that the low-energy theory is a two-Higgs-doublet 
model. In this framework, one of Higgs doublets, which gives mass to the 
up-type quarks, is assumed to arise from the ${\bf 15}$ of $SU(6)$. This 
allows all the up-type quark masses to be generated from renormalizable 
terms of the form $\epsilon^{\kappa \lambda \mu \nu \rho \sigma} 
{\bf{15_{\rm H}}}_{\kappa \lambda} {\bf {15}}_{\mu \nu} {\bf {15}}_{\rho 
\sigma}$, where the Higgs doublet is now contained in the ${\bf 15_{\rm 
H}}$~\cite{Kim:1981jw}. The other Higgs doublet, which arises from the 
${\bf 6}$ of $SU(6)$, gives mass to the down-type quarks and charged 
leptons. The central observation is that the same Higgs doublet in the 
${\bf 15_{\rm H}}$ that generates the large top quark mass can also be 
used to generate a Dirac neutrino mass term through renormalizable 
operators of the form ${y_\nu}^{ij} {\bf 15_{\rm H}}_{\mu \lambda} {\bf 
\bar{6}}^{\mu}_i {\bf \bar{6}}^{\lambda}_j$, where $i$ and $j$ are 
flavor indices. Since the ${\bf 15}$ of $SU(6)$ is antisymmetric in its 
tensor indices, this vanishes if the flavor indices $i$ and $j$ are the 
same. Therefore, this construction naturally leads to a skew-symmetric 
structure for the Dirac mass matrix of the neutrinos in flavor space.

This framework can naturally accommodate either Dirac or Majorana 
neutrino masses. The right-handed neutrinos can naturally acquire large 
Majorana masses of order $M_{\rm GUT}^2/M_{\rm Pl} \sim 10^{14}$ GeV 
from nonrenormalizable Planck-suppressed interactions with the Higgs 
fields that break the GUT symmetry. This naturally leads to Majorana 
masses for the neutrinos of the right size through the seesaw 
mechanism~\cite{Minkowski:1977sc, Mohapatra:1979ia, Yanagida:1979as, 
GellMann:1980vs}. Alternatively, as a consequence of additional discrete 
symmetries, a Majorana mass term for the right-handed neutrinos may not 
be allowed, while the coefficient of the Dirac mass term is suppressed. 
In such a scenario we obtain Dirac neutrino masses. In this paper we will 
consider both the Dirac and Majorana cases.

This paper is organized as follows. In Section~\ref{sec:II}, we outline 
the framework that underlies this class of models and show how the 
pattern of neutrino masses emerges in the Dirac and Majorana cases. In 
Section~\ref{numassmat}, we present a realistic model in which the 
neutrino masses are Dirac, and perform a detailed numerical fit to the 
neutrino masses and mixings using a recent global analysis of the 
3-neutrino oscillation data. We show that this framework predicts an 
inverted spectrum of neutrino masses with one mass eigenstate 
hierarchically lighter than the others. In Section~\ref{sec:Maj_option}, 
we present a realistic model in which the neutrino masses are Majorana, 
and again perform a detailed numerical fit to the neutrino oscillation 
data. We show that in this scenario one neutrino is again hierarchically 
lighter than the others, but the spectrum of neutrino masses can now be 
either normal or inverted. We also explore the implications of this 
scenario for future $0\nu\beta\beta$ experiments and future cosmological 
observations. Our conclusions are presented in Section~\ref{sec:V}.

\section{The Framework} \label{sec:II}

Our model is based on the $SU(6)$ GUT symmetry with the fermions of each 
family arising from a ${\bar {\bf 6}}$ representation, denoted by 
${\chi}$, and a rank-two antisymmetric representation ${\bf 15}$, 
denoted by $\psi$. For now we omit the generation indices. Note that 
anomaly cancellation for the $SU(6)$ group requires that there be two 
${\bar {\bf 6}}$ chiral fermion representations for each {\bf 15} 
fermion. We denote the additional ${\bar {\bf 6}}$ of each family by 
$\hat{\chi}$. After the breaking of $SU(6)$ to $SU(5)$, the fields in 
$\hat{\chi}$ that carry charges under the SM gauge groups acquire large 
masses at the GUT scale by marrying the non-SM fermions in the ${\bf 
15}$. Therefore, these fields do not play a role in generating the 
masses of the light fermions. However, the SM-singlet field in 
$\hat{\chi}$, which has no counterpart in the ${\bf 15}$, may remain 
light. We employ the familiar convention in which all fermions are taken 
to be left-handed, and the SM fermions are labelled as $(Q, u^c, d^c, L, 
e^c)$, with $Q^T = (u, d)$ and $L^T = (\nu, \ell)$.

The $SU(6)$ symmetry is broken near the GUT scale down to $SU(5)$, which 
contains the usual embedding of SM fermions in a ${\bf \bar{5}}$ and a 
{\bf 10} of $SU(5)$. Without loss of generality we take the $SU(5)$ 
indices to be $(2,3,4,5,6)$, so that the index $1$ lies outside $SU(5)$. 
Color indices run over $(4,5,6)$. 

We now consider the assignment of fermions under representations of 
$SU(6)$. Under the fermion multiplet $\chi$ that transforms as a ${\bf 
\overline{6}}$, we have
 \begin{align}
  \chi
  & \ = \ 
  \left(
  \begin{array}{c}
    \nu^c \\
    \hline
    L\\
    \hline
    d^c \\
  \end{array}
  \right) \; ,
 \end{align} 
 where $L$ is the SM lepton doublet, $L^T = (\nu, \ell)$. Note that the 
Dirac partner $\nu^c$ of the SM neutrino is embedded in the same 
multiplet as the left-handed leptons. The fermions in $\hat{\chi}$ also 
transform as $\bar{\bf 6}$:
 \begin{align} 
 \hat{\chi}        
  & \ = \ 
  \left(
  \begin{array}{c}
    {N}^c \\
    \hline
    \hat{L}\\
    \hline
    \hat{D}^c \\
  \end{array}
  \right) \; .
 \end{align}
 The fermion content of $\psi$, which transforms as a {\bf 15}-dimensional 
representation of $SU(6)$, is given by
 \begin{align}
  \psi
  &
 \ = \ 
  \left(
  \begin{array}{c|cc|ccc}
    0 & \multicolumn{2}{c|}{\hat{L}^c} & \multicolumn{3}{c}{ \hat{D} }\\
\hline
      &  0 & e^c  & \multicolumn{3}{c}{d} \\
      &   &  0   & \multicolumn{3}{c}{u} \\
\hline
      &   &   &  0 & u^c_3  & -u^c_2 \\
      &   &   &   & 0  & u^c_1 \\
      &   &   &   &   & 0\\
  \end{array}
  \right) \, .
\end{align}

 The breaking of $SU(6)$ down to $SU(5)$ at the GUT scale is realized by a 
Higgs field $\hat{H}$ which transforms as a ${\bf 6}$ under $SU(6)$ and 
acquires a large vacuum expectation value (VEV) along the SM-singlet 
direction. A Higgs field $\hat{\Sigma}$, which transforms as an adjoint 
under $SU(6)$, further breaks $SU(5)$ down to the SM gauge group. The 
breaking of electroweak symmetry is realized through two Higgs doublets 
$H$ and $\Delta$ that arise from different $SU(6)$ representations. The 
field $H$, which gives masses to the down-type quarks and charged 
leptons, emerges from a {\bf 6} while $\Delta$, which gives masses to 
the up-type quarks, arises from a {\bf 15}. The Higgs fields $\hat{H}$, 
$H$ and $\Delta$ are assumed to have the following VEVs:
 \begin{align}
  \langle \hat{H} \rangle
  & \ = \ 
  \left(
  \begin{array}{c}
      M \\
      0 \\
      0 \\
      0\\
      0\\
      0
  \end{array}
  \right) \, , 
  &
  \langle H \rangle
  & \ = \ 
  \left(
  \begin{array}{c}
      0 \\
      v_d \\
      0 \\
      0\\
      0\\
      0
  \end{array}
  \right) \, , 
  &
  \langle \Delta \rangle
  & \ = \ 
  \left(
  \begin{array}{cccccc}
    0 & v_u & 0 & 0 & 0 & 0 \\
   -v_u & 0 & 0 & 0 & 0 & 0 \\
    0 & 0 & 0 & 0 & 0 & 0 \\
    0 & 0 & 0 & 0 & 0 & 0\\
    0 & 0 & 0 & 0 & 0 & 0\\
    0 & 0 & 0 & 0 & 0 & 0
  \end{array}
  \right)
\; .
 \end{align}
 The VEV of $\hat{\Sigma}$ takes the pattern 
 \begin{align}
 \label{eq:VEV-Sigma}
 \langle \hat{\Sigma} \rangle
  & \ = \  \hat{M}
  \left(
  \begin{array}{cccccc}
    0 & 0 & 0 & 0 & 0 & 0 \\
    0 & -\frac{3}{2}& 0 & 0 & 0 & 0 \\
    0 & 0 & -\frac{3}{2} & 0 & 0 & 0 \\
    0 & 0 & 0 & 1& 0 & 0\\
    0 & 0 & 0 & 0 & 1 & 0\\
    0 & 0 & 0 & 0 & 0 & 1    
  \end{array}
  \right)
\; .
 \end{align}
 The field content is summarized in Table~\ref{tab:field}. Here $N_F$ denotes the number of flavors. 
 \begin{table}[t!]
  \begin{tabular}{||c|c|c|c||}
    \hline \hline
 \multicolumn{2}{||c|}{Multiplets}   & $SU(6)$ representation & $N_F$ \\
    \hline
& $\chi$ & $\overline{\bf 6}$  & 3 \\
 fermion  & $\hat{\chi} $& $\overline{\bf 6}$  & 3 \\
 &   $ \psi$ & {\bf 15}  & 3\\ \hline
 &   $H$ & $\overline{\bf 6}$ & 1\\
scalar  &    $\hat{H}$ & $\overline{\bf 6}$ & 1\\
  &  $\Delta$ & $\overline{\bf 15}$ & 1\\
  &  $\hat{\Sigma}$ & {\bf 35} & 1\\
    \hline\hline
  \end{tabular}
  \caption{Field content of the $SU(6)$ model under consideration.} 
\label{tab:field} \end{table}

We now discuss the generation of fermion masses. The additional 
fermions $\hat{L}$, $\hat{D}^c$ in $\hat{\chi}$ and $\hat{L}^c$,  
$\hat{D}$ in $\psi$ acquire masses at the GUT scale through interactions 
with $\hat{H}$ of the form 
 \begin{align}
\label{decouple}
- {\cal L}_{\rm decouple}  \ = \ \hat{\lambda}_{i j} \psi_i \hat{\chi}_{j} \hat{H}
  +\mathrm{h.c.} \;,
 \end{align}
 where we have suppressed the $SU(6)$ and Lorentz indices and shown only 
the flavor indices. Consequently, these fields do not play any role in 
the generation of the masses of the SM fermions. These interactions do 
not give mass to the SM-singlet field $N^c$ in $\hat{\chi}$. However, 
even if $N^c$ is light, the fact that it is a SM singlet means that in 
the absence of other interactions its couplings to the SM fields at low 
energies are very small.

The SM fermions acquire masses from their Yukawa couplings to the Higgs 
fields $H$ and $\Delta$ after electroweak symmetry breaking. The 
$SU(6)$-invariant Yukawa couplings take the form
 \begin{align}
 \label{SMfermionmass}
- {\cal L}_Y  \ = \ y_{d,ij} \psi_i \chi_j H + y_{u,ij} \psi_i \psi_j {\Delta^\dagger}
  +\mathrm{h.c.}
 \end{align}
 The down-quark and charged-lepton masses arise from the $y_d$ term in 
the Lagrangian after the Higgs field $H$ acquires an electroweak-scale 
VEV. Similarly the up-quark masses arise from the $y_u$ term in the 
Lagrangian after $\Delta$ acquires a VEV. In general, the masses of the 
SM fermions also receive contributions from higher-dimensional operators 
suppressed by the Planck scale ($M_{\rm Pl}$) that involve 
$\hat{\Sigma}$, such as
 \begin{equation}
 \label{SMfermionmassCorrections}
- {\cal L}_{\Delta Y}  \ = \  \frac{\hat{y}_{d,ij}}{M_{\rm Pl}} \psi_i \chi_j \hat{\Sigma} H 
+ \frac{\hat{y}_{u,ij}}{M_{\rm Pl}} \psi_i \psi_j \hat{\Sigma}{\Delta^\dagger}
+ \mathrm{h.c.}
 \end{equation}
 The VEV of $\hat{\Sigma}$ breaks the $SU(5)$ symmetry that relates 
quarks and leptons [cf.~Eq.~\eqref{eq:VEV-Sigma}]. Therefore these 
higher-dimensional operators violate the GUT symmetries that relate the 
masses of the down-type quarks to those of the leptons of the same 
generation.

A Dirac mass term for the neutrinos may be obtained from interactions of 
the form
 \begin{align}
- {\cal L}_{D}  \ = \  y_{\nu,ij} \chi_i \chi_j \Delta^\dagger +\mathrm{h.c.}
\label{Dmass}
 \end{align}
 As explained earlier, the fact that $\Delta$ is an antisymmetric tensor 
under $SU(6)$ implies that $y_{\nu,ij}$ is skew-symmetric in flavor space. 
Consequently, the resulting Dirac mass matrix for the neutrinos has 
vanishing determinant. We expect corrections to the Dirac mass term from
Planck-suppressed higher-dimensional operators, such as
 \begin{align}
-  \mathcal{L}_{\Delta D}  \ = \ 
   \frac{\kappa_{\nu,ij}}{M_{\rm Pl}}\chi_i H^\dagger \chi_j \hat{H}^\dagger
  +{\rm h.c.}
 \end{align} 
 In general, this contribution will be suppressed by a factor $M_{\rm 
GUT}/M_{\rm Pl} \sim 10^{-2}$ relative to that from Eq.~(\ref{Dmass}).

A large Majorana mass term for the right-handed neutrinos can be obtained
from Planck-suppressed nonrenormalizable interactions of the form
 \begin{align}
- \mathcal{L}_{M}  \ = \ \frac{\lambda_{\nu^c, ij}}{M_{\rm Pl}} \hat{H} ^\dagger\chi_i \hat{H}^\dagger \chi_j \; .
\label{Mmass}
 \end{align}
 This leads to Majorana masses for the right-handed neutrinos of order 
$M_{\rm GUT}^2/M_{\rm Pl}$, which is parametrically of order the seesaw 
scale $\sim 10^{14}$ GeV. Then, from Eqs.~(\ref{Dmass}) and (\ref{Mmass}), we 
obtain Majorana neutrino masses of the right size.

If neutrinos are to be Dirac particles, the mass term for the 
right-handed neutrinos shown in Eq.~(\ref{Mmass}) must be absent. 
Furthermore, we require the coefficients of the Dirac mass terms to be 
extremely small, $y_{\nu,ij}, \kappa_{\nu,ij} \sim 10^{-11}$, to 
reproduce the observed values of the neutrino masses. In Section~\ref{numassmat}, we 
shall show that the absence of the Majorana mass term for the 
right-handed neutrinos, Eq.~(\ref{Mmass}), and the smallness of 
$y_{\nu,ij}$ and $\kappa_{\nu,ij}$ can be explained on the basis of 
discrete symmetries.

\section{Dirac Neutrino Masses}\label{numassmat}

\subsection{Pattern of Neutrino Masses} \label{sec:IIIA}

We now present a simple model that realizes the pattern of Dirac 
neutrino masses discussed in Section~\ref{sec:II}.  The model is based on 
discrete $Z_4\times Z_7$ symmetries under which the fermions and Higgs 
scalars have the charge assignments shown in Table~\ref{Diraccharges}.
 \begin{table}
 \begin{tabular}{||c|c|c|c|c||} \hline\hline
 \multicolumn{2}{||c|}{Multiplets}  & $SU(6)$ representation & $Z_4$ quantum number & $Z_7$ quantum number \\ \hline
 & $\chi$ & ${\bf \bar{6}}$ & $+1$  & $+4$ \\
fermion &  $\hat{\chi}$ & ${\bf\bar{ 6}}$ & $-1$ & $-1$ \\
&  ${\psi}$ & ${\bf {15}}$ & $+1$ & $+1$ \\ \hline
& {$H$} & $\overline{\bf 6}$ & $+2$ & $+2$ \\
& ${\hat{H}}$ & $\bar {\bf 6}$ & $0$ & $0$ \\
scalar & $\Delta$ & $\overline{\bf 15}$ & $+2$  & $+2$ \\
& $\hat{\Sigma}$ & {\bf 35} & $0$ & $0$ \\
& $\sigma$ & {\bf 1} & $0$ & $+1$ \\
\hline
 \end{tabular}
 \caption{Quantum numbers of the various fermion and scalar fields under 
the discrete $Z_4\times Z_{7}$ symmetry in the model of Dirac neutrinos. 
Here the integer entries $n$ correspond to transformation under $Z_4$ as 
$e^{2\pi i n/4}$ and under $Z_7$ as $e^{2\pi i n/7}$.}
 \label{Diraccharges}
 \end{table}
 The Yukawa couplings that generate masses for the SM fermions, 
Eqs.~(\ref{SMfermionmass}) and (\ref{SMfermionmassCorrections}), are 
consistent with the $Z_4$ and $Z_7$ symmetries. The interaction in 
Eq.~(\ref{decouple}) that gives GUT-scale masses to the extra fermions 
$\hat{L}$, $\hat{D}^c$ in $\hat{\chi}$ and $\hat{L}^c$, $\hat{D}$ 
in $\psi$ is also allowed by the discrete symmetries. However, the 
renormalizable Dirac mass term for the neutrinos, Eq.~(\ref{Dmass}), is 
now forbidden by the discrete $Z_7$ symmetry. Instead, the leading 
contribution to the neutrino masses arises from the dimension-5 term
 \begin{align}
- \mathcal{L}_{d=5}  \ = \  y_{\nu,ij} \frac{\sigma}{M_{\rm Pl}} \chi_i \chi_j \Delta^\dagger +\mathrm{h.c.}
 \label{nuDmass}
 \end{align}
 The field $\sigma$, which is a singlet under $SU(6)$, is assumed to acquire a VEV, thereby spontaneously 
breaking the discrete $Z_7$ symmetry.  For $\langle \sigma \rangle \sim 
10^{7}$ GeV, we obtain Dirac neutrino masses in the right range. Since 
$\Delta$ is in an antisymmetric representation of $SU(6)$, these mass 
terms are antisymmetric in flavor space, i.e. 
 \begin{align}
  y_{\nu,ij} \ = \ - y_{\nu,ji} \ .
 \end{align} 
 This leads to a highly predictive spectrum, with one zero eigenvalue, 
and the other two eigenvalues equal in magnitude and opposite in sign. 
This corresponds to an inverted mass hierarchy, in which the smaller $\Delta 
m^2$ arises from the difference between the masses of the two heavier 
eigenstates. We can perform phase rotations on the right-handed 
neutrinos to ensure that the elements of this mass matrix are real, so 
that the phase in the PMNS matrix vanishes.

 Clearly, the mass pattern above is ruled out experimentally. However, 
we need to include the effects of higher-dimensional terms, which will 
give corrections to the pattern above. Since these corrections are 
expected to be small, we expect to retain the qualitative features of 
the spectrum above, in particular, an inverted ordering. An example of 
such a higher-dimensional operator is the dimension-6 term
 \begin{align}
 - \mathcal{L}_{\rm d=6}
  & \ = \ 
  \kappa_{\nu,ij}
  \frac{\sigma}{M_{\rm Pl}^2}\chi_i H^\dagger \chi_j \hat{H}^\dagger
  + {\rm h.c.}
 \end{align}
 This correction is parametrically smaller than the antisymmetric 
contribution in Eq.~(\ref{nuDmass}) by a factor $M_{\rm GUT}/M_{\rm Pl} 
\sim 10^{-2}$. 

In order for the terms in Eq.~(\ref{nuDmass}) to give rise to the 
leading contribution to the neutrino masses, other possible mass terms 
involving the light neutrino fields $\nu$ and $\nu^c$ must be 
suppressed.  The discrete $Z_4$ symmetry forbids Majorana mass terms 
for $\nu$ and $\nu^c$. It also forbids Dirac mass terms 
between $\nu$ and $N^c$. A Dirac mass term between $\nu^c$ and $N^c$ can 
be generated as a $Z_7$-breaking effect, but only at dimension-8:
  \begin{align}
- \mathcal{L}_{d=8} \ = \  \frac{{\sigma^\dagger}^3}{M_{\rm Pl}^4} \hat{\chi} \hat{H}^\dagger \chi \hat{H}^\dagger
  + {\rm h.c.}
 \end{align}
 This is too small to have any observable effect. Therefore, without 
loss of generality, the neutrino mass matrix has the form of a real 
skew-symmetric matrix with a small complex symmetric component. We 
write the mass term in matrix form as,
 \begin{align}
-  \mathcal{L}_{\rm mass}
   \ = \ 
  \left(
  \begin{array}{ccc}
    \nu^c_e & \nu^c_\mu & \nu^c_\tau 
  \end{array}
  \right) M_\nu
  \left(
  \begin{array}{c}
    \nu_e \\
    \nu_\mu \\
    \nu_\tau
  \end{array}
  \right)  \;.
\label{eq:diracM}
 \end{align}
 It is convenient to decompose the Dirac mass matrix as,
 \begin{equation}
 \label{eq:mass1}
  M_\nu \ = \ M_\nu^0 + \delta m \;.
 \end{equation} 
 Here $M_\nu^0$ is skew-symmetric and takes the form 
 \begin{align}
 M_\nu^0 \ = \  \left(
  \begin{array}{ccc}    0 & m_a & m_b \\
    -m_a & 0 & m_c \\
    -m_b & -m_c & 0 
  \end{array}
  \right) \ ,
  \label{eq:diracM0}
 \end{align}
 while $\delta m$ is an anarchic symmetric matrix whose entries are 
parametrically smaller than those in $M_\nu^0$. We can choose $m_a, m_b$ 
and $m_c$ in Eq.~\eqref{eq:diracM0} to be real without loss of 
generality. However, in general the elements of $\delta m$ are complex.

The PMNS matrix $U$ is, as usual, defined to be the rotation matrix that 
relates the flavor eigenstates $\nu_\ell$ of the active neutrinos to the mass 
eigenstates $\nu_i$:
 \begin{align}
\left(
\begin{array}{c}
\nu_e  \\
\nu_\mu \\
\nu_\tau
\end{array}
\right) 
& \ = \ 
U
\left(
\begin{array}{c}
\nu_1  \\
\nu_2 \\
\nu_3
\end{array}
\right) \, .
 \end{align}
 Defining $D_\nu={\rm diag}(m_1,m_2,m_3)$ as the diagonalized mass 
matrix with mass eigenvalues $m_i$ corresponding to the eigenstates 
$\nu_i$, we have
 \begin{align}
 D_\nu^\dagger D_\nu \ = \ U^\dagger M_\nu^\dagger M_\nu U \, .
 \end{align}
 Therefore the PMNS matrix is identified with the matrix that 
diagonalizes the matrix $M_\nu^\dagger M_\nu$. By a suitable choice of 
of $m_a, m_b, m_c$, and the elements in $\delta m$, we can fit the 
observed neutrino mass splittings and mixing angles.

 Before proceeding with a numerical scan, we first estimate the region 
of parameter space consistent with observations. Although there are a 
large number of free parameters, since only $m_a, m_b$ and $m_c$ are 
expected to be large, this scenario is very predictive. We parametrize 
the elements of the skew-symmetric matrix $M_\nu^0$ as follows:
 \begin{eqnarray}
m_a & \ = \ & m \cos\theta \cos\phi \, , \nonumber \\
m_b & \ = \ & m \cos\theta \sin\phi \, , \nonumber \\
m_c & \ = \ & m \sin\theta \, . 
\label{eq:DiracParam}
 \end{eqnarray}
 Since $\delta m$ arises from a higher-dimensional operator, it can be 
treated as a perturbation. At zeroth order in this perturbation, the 
eigenvalues for $M_\nu^\dagger M_\nu$ are simply $\{m^2, m^2, 0\}$. This 
corresponds to a limiting case of an inverted mass hierarchy in which 
the smaller (solar) mass splitting vanishes. By convention, in an inverted 
hierarchy the mass eigenstates $m_1, m_2, m_3$ are labeled such that 
$m_3$ corresponds to the mass of the lightest state and the smaller 
splitting is between $m_1$ and $m_2$, with $m_2 > m_1$. In our case, 
these correspond to the masses of two degenerate eigenstates with mass 
$m$. Then the eigenstate with vanishing mass is identified as $\nu_3$. 
The mixing angle $\theta_{12}$ mixes states in the degenerate subspace, 
and hence is arbitrary at this order. It will be fixed by the 
perturbation. The other two mixing angles are given by $\theta_{13} = \theta$ 
and $\theta_{23} = \phi$. The Dirac $C\!P$  phase $\delta_{C\!P}$ can be rotated away at 
this order as well.

To summarize, for $\delta m=0$, which corresponds to zeroth order in the 
perturbation, the model predictions for the solar and atmospheric 
mass-squared splittings, the mixing angles, and the Dirac $C\!P$ phase
are given by
 \begin{align}
 & \Delta m^2_{\rm sol}  \ \equiv \ \Delta m^2_{21} \ = \ 0 \, , \quad 
  \Delta m^2_{\rm atm}  \ \equiv \ |\Delta m^2_{32}| \ = \    m^2 \, , \nonumber \\
&  \theta_{13}  \ = \ \theta \, ,\quad 
  \theta_{23} \ = \ \phi \, , \quad 
\theta_{12}  \ = \ \mathrm{arbitrary} \, , \quad 
  \delta_{C\!P} \ = \ 0 \; ,
  \label{eq:23}
\end{align}
where $\Delta m^2_{ij}\equiv m_i^2-m_j^2$. Once we add the perturbation $\delta m$, the solar splitting and the mixing angle 
$\theta_{12}$ are fixed. The perturbation $\delta m$ can be parametrized 
as $\eta\, \widehat m$, where $\hat{m}$ is an anarchic symmetric matrix 
with entries of order $m$. The lightest eigenstate acquires a mass of 
order $\eta m$ from the perturbation, and the solar splitting is now 
\begin{align}
\Delta m^2_{\rm sol} \ \equiv   \ m_2^2-m_1^2  \ \sim \ 2 \eta m^2 \, . 
\end{align}
The atmospheric mass 
splitting $\Delta m^2_{\rm atm}\equiv |m_{3}^2-m_2^2|$ continues to remain of the order of $m^2$. The 
ratio of the solar and atmospheric splittings determines the parametric 
size of $\eta$, which in turn sets the mass of the lightest eigenstate. 
Putting in the numbers, we have
 \begin{align}
  m_1& \ \simeq \ \sqrt{\Delta m^2_{\rm atm}}  \ \sim \ 0.05\  \mathrm{ eV} \, , \nonumber \\
  m_2& \ \simeq \  m_1 + \frac{\Delta m^2_{\rm sol}}{2m_1} \ 
  \sim  \ 0.05\ \mathrm{eV} \, , \nonumber \\
  m_3& \ \simeq \  \frac{\Delta m^2_{\rm sol}}{2\sqrt{\Delta m^2_{\rm atm}}}
 \  \sim \ 7 \times 10^{-4} \ \mathrm{
  eV} \, .
 \end{align}
 We see that a satisfactory fit to the data requires the parameter 
$\eta$ to be of order $m_3/m_1 \sim 10^{-2}$. Remarkably, this is in 
excellent agreement with the expected value of $\eta$ from our 
construction, $\eta \sim M_{\rm GUT}/M_{\rm Pl} \sim 10^{-2}$.
 
We see that this flavor pattern results in a very predictive spectrum of 
neutrino masses and mixings. We obtain an inverted mass hierarchy, with 
one neutrino hierarchically lighter than the other two. This prediction 
can be conclusively tested in future long-baseline oscillation 
experiments such as Hyper-K~\cite{Abe:2018uyc} and 
DUNE~\cite{Abi:2018dnh}. Since the $C\!P$-violating phase 
$\delta_{C\!P}$ in the PMNS matrix vanishes in the limit that $\delta m$ 
is zero, it might have been expected to be small. However, the results 
of our numerical scans in Section~\ref{subsec:fit1} show that this need 
not be the case, and that fairly large values of $\delta_{C\!P}$ can be 
obtained even for $\eta \lesssim 10^{-2}$.

\begin{table}[t!]
\centering
\begin{tabular}{||c|c|c|c|c|c|c|c|c|c||}
\hline \hline
Fit  & $|x_{11}|$ & $|x_{22}|$ & $x_{33}$ & $x_{12}$ & $x_{13}$ & $x_{23}$ & $\varphi_{11}$ & $\varphi_{22}$ \\ \hline \hline
Fit 1 (IH)  & 0.0620 & 0.0180 & 0.0410 & 0.0088 & 0.0184 & 0.0075 & $227.18^\circ$  & - \\ \hline
Fit 2 (IH)  & 0.1012 & 0.0234 & 0.0202 & 0.0113 & 0.0151 & 0.0022 & $292.30^\circ$ & - \\ \hline
Fit 3 (IH)  & 0.0620 & 0.0604 & 0.0239 & 0.0038 & 0.0236 & 0.0041 & $269.50^\circ$ & $288.10^\circ$  \\
     \hline \hline     
\end{tabular}
\vspace{0.1cm}
 \caption{The values of the parameters for three benchmark points chosen 
to fit the neutrino oscillation data in the case of Dirac neutrinos. }
 \label{nuTab1}
\vspace{1cm}
\begin{tabular}{||c|c|c|c|c||}
\hline \hline
 {Oscillation}   &   {{$3\sigma$ allowed range}} &  \multicolumn{3}{c||}{{Model prediction}}\\
 \cline{3-5}
 {parameters} & {from {\tt NuFit4.1}}~\cite{Esteban:2018azc} & Fit 1 (IH) &  Fit 2 (IH) & Fit 3 (IH)  \\ \hline \hline
 $\Delta m_{21}^2 (10^{-5}~{\rm eV}^2$)  &   6.79 - 8.01   &  7.35 & 7.39  & 7.41\\ \hline
 $\Delta m_{23}^2 (10^{-3}~{\rm eV}^2) $ &   2.416 - 2.603  & 2.540 &  2.506 & 2.540 \\ \hline
  $\sin^2{\theta_{12}}$   &   0.275 - 0.350 &   0.319 &  0.314 & 0.305 \\ \hline
 $\sin^2{\theta_{23}}$   &   0.430 - 0.612  &   0.557 &  0.558 & 0.559 \\ \hline
  $\sin^2{\theta_{13}} $   &   0.02066 - 0.02461  &   0.0230 &  0.0224 & 0.0227 \\ \hline
  $\delta_{C\!P}~(^\circ)$ & 205 - 354 & 330.8 & 277.7  & 287.7
  \\ \hline 
 $m_3$ ($10^{-4}~{\rm eV}$) & - & $1.57$ & $1.56$ & $2.88$ \\ \hline
  \hline
\end{tabular}
 \caption{
 Predictions of the three benchmark points for the neutrino oscillation 
parameters in the case of Dirac neutrinos, compared to the 3$\sigma$ 
allowed range from a recent global fit. Also included are the 
predictions of the benchmark points for the mass of the lightest 
neutrino.
 }
 \label{tab:IHDirac}
\end{table}
\begin{figure}[t]
$$\includegraphics[scale=0.3]{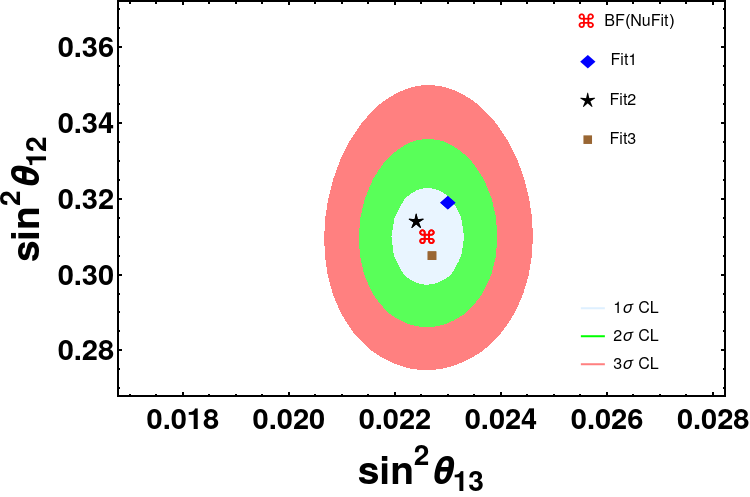} \hspace{3mm}
\includegraphics[scale=0.3]{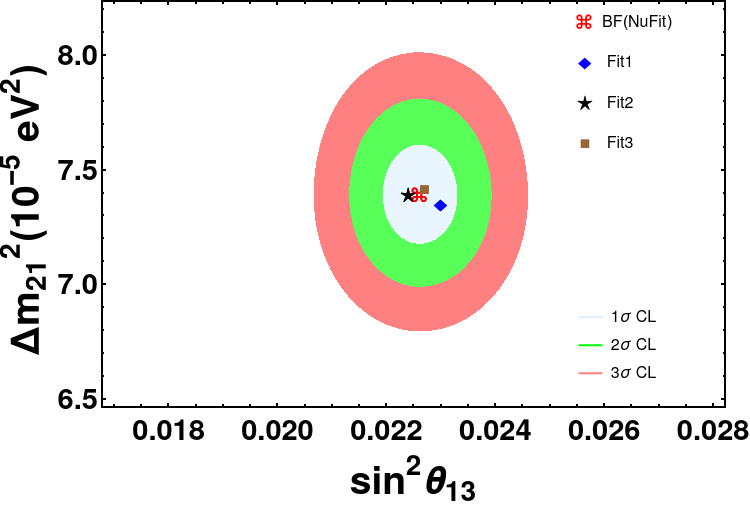} $$
$$ \includegraphics[scale=0.3]{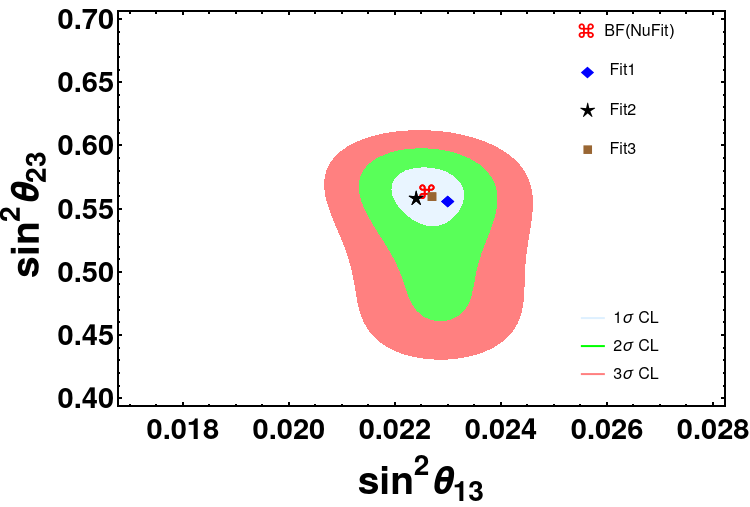} \hspace{3mm}
\includegraphics[scale=0.3]{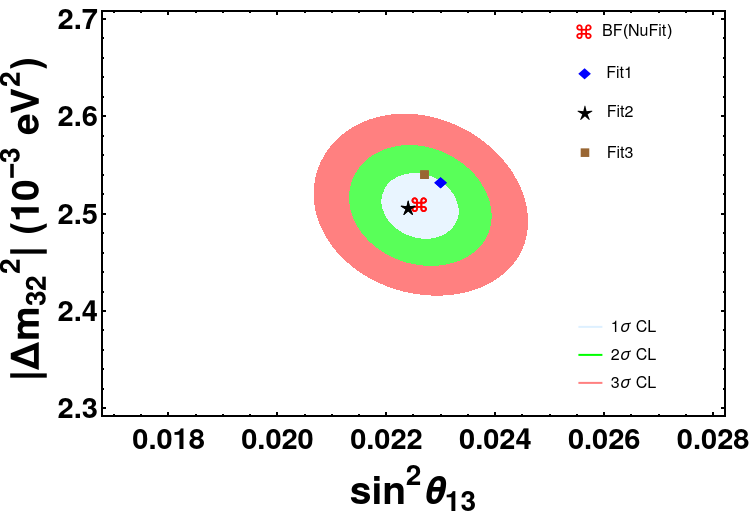}  $$
 \caption{Global oscillation analysis obtained from {\tt NuFit4.1} for 
the case of an inverted hierarchy (IH) compared to the results from our 
benchmark points for the Dirac model (Fit1, Fit2, Fit3). The gray, green, and 
pink-colored contours represent the NuFit $1\sigma, 2\sigma, \text{and} 
\, 3\sigma $ CL allowed regions respectively, while the red markers 
represent the NuFit best-fit values for an IH. The blue, black, and brown markers 
are respectively the predictions of the benchmark points corresponding to Fit 1,
Fit 2, and Fit 3, as given in Table \ref{tab:IHDirac}. }
 \label{fig:nufit_dirac}
\end{figure}
\subsection{Fits to the Data}\label{subsec:fit1}

Our strategy for the scan is as follows. The neutrino mass matrix is 
parameterized in terms of a skew-symmetric matrix $M_\nu^0$ with a small 
symmetric correction $\delta m$, as discussed in Section~\ref{sec:IIIA}. 
We fix the parameters $\{m_a,m_b,m_c\}$ of the skew-symmetric matrix 
$M_\nu^0$ in Eq.~\eqref{eq:diracM0} such that the zeroth order 
predictions match the measured values of $\Delta m^2_{\rm atm}$, 
$\theta_{13}$ and $\theta_{23}$ as given by Eq.~\eqref{eq:23}. In 
particular, we take $m^2 \equiv \Delta m_{\rm atm}^2 = 2.509 \times 
10^{-3}$ eV$^2$, $\theta \equiv \theta_{13} = 8.61^\circ$, and $\phi 
\equiv \theta_{23}= 48.3^\circ$ corresponding to the central values from 
NuFit~\cite{Esteban:2018azc} for the inverted hierarchy case and employ 
Eq.~(\ref{eq:DiracParam}) to determine $m_a, m_b$, and $m_c$. Further, 
the size of the perturbation $\eta$ is fixed by $\Delta m^2_{\rm sol} / \Delta m_{\rm atm}^2$. 
We then scan over the anarchic matrix $\widehat{m}$ and obtain numerical 
predictions for the entire PMNS matrix. We choose to parametrize the 
mass matrix in Eq.~\eqref{eq:mass1} in terms of $m_c$ and the ratios 
$x_1 \equiv {m_a}/{m_c}$, $x_2 \equiv {m_b}/{m_c}$ and $x_{ij} \equiv 
{\delta m_{ij}}/{m_c}$,
 \begin{align}
  M_\nu
  =
  \left(
  \begin{array}{ccc}
    0 & m_a & m_b \\
    -m_a & 0 & m_c \\
    -m_b & -m_c & 0 
  \end{array}
  \right) + \delta m
   &= m_c \left[ \left(
  \begin{array}{ccc}
    0 & x_1 & x_2 \\
    -x_1 & 0 & 1 \\
    -x_2 & -1 & 0 
  \end{array}
 \right) + \left(
  \begin{array}{ccc}
    |x_{11}|e^{i\varphi_{11}} & x_{12} & x_{13} \\
    x_{12} & |x_{22}|e^{i\varphi_{22}}& x_{23} \\
    x_{13} & x_{23} & x_{33}
  \end{array}
  \right) \right] \ . 
  \end{align} 
 As can be seen from Eq.~(\ref{eq:DiracParam}), the values of $x_1$ and 
$x_2$ are fixed at 4.393 and 4.931, respectively. The elements of the 
perturbation matrix $\delta m$ are restricted to be much smaller than 
$m_a$, $m_b$, and $m_c$. The input parameters $x_{ij}$ shown in Table 
\ref{nuTab1} are examples of fits that are in excellent agreement with 
the recent global fit results from NuFit~\cite{Esteban:2018azc}. In 
obtaining these fits, all the elements of $\delta m$ have been taken to 
be real except $\delta m_{11}$ and $\delta m_{22}$. We have introduced 
phases $\varphi_{11}$ and $\varphi_{22}$ in the elements $\delta m_{11}$ 
and $\delta m_{22}$ respectively in order to obtain a non-zero $C\!P$ 
phase in the PMNS matrix. Although the addition of just a single phase, 
say $\varphi_{11}$, can give us a non-vanishing $\delta_{C\!P}$ (as in 
Fits 1 and 2), we find that in this case a large $\delta_{C\!P}$ 
requires a somewhat larger value of $|x_{11}|$ (as in Fit 2). The 
addition of a second phase $\varphi_{22}$ allows us to obtain a large 
$\delta_{C\!P}$ even if all the $x_{ij}$ are small (as in Fit 3).

The predictions of these fits for the oscillation parameters are shown 
in Table \ref{tab:IHDirac}, along with the $3 \sigma$ allowed range from 
NuFit4.1 global analysis~\cite{Esteban:2018azc}. Also included are the 
predictions for the mass of the lightest neutrino. Note that in each of 
these fits the lightest neutrino mass is hierarchically lighter than the 
other two mass eigenstates by more than two orders of magnitude. The 
results for the fits presented in Table~\ref{tab:IHDirac} are also 
displayed in Fig.~\ref{fig:nufit_dirac} as Fit1, Fit2 and Fit3 in a 
two-dimensional projection of the $1\sigma$ (gray), $2\sigma$ (green), 
and $3\sigma$ (pink) confidence level (CL) regions of the global-fit 
results (without the inclusion of the Super-K atmospheric $\Delta 
\chi^2$-data). The NuFit best-fit points in each plane are shown by the 
red markers, while the blue, black and brown markers correspond to Fit1, 
Fit2 and Fit3 respectively.

Interestingly, we find no significant restriction on the 
$C\!P$-violating phase $\delta_{C\!P}$ in the PMNS matrix in this 
scenario. In particular, as seen from Fit 3, we can get a large $C\!P$ 
phase in the PMNS matrix even if all the elements of $\delta m$ are 
smaller by a factor of order $10^{-2}$ than the observed atmospheric 
splitting. Larger $\delta_{C\!P}$ values seem to be preferred by the 
recent T2K results~\cite{Abe:2019vii}, and in the future, a more precise 
determination of $\delta_{C\!P}$ can only help us better constrain the 
parameter space of the model. 


\section{Majorana Neutrino Masses} \label{sec:Maj_option}

\subsection{Pattern of Neutrino Masses}

We now present a simple model in which the pattern of Majorana neutrino 
masses discussed in Section~\ref{sec:II} is realized.  The model is based on a 
discrete $Z_6$ symmetry under which the fermions and Higgs scalars have 
the charge assignments shown in Table~\ref{tab:II}.
\begin{table}[t!]
 \begin{center}
\begin{tabular}{||c|c|c|c||} \hline\hline
\multicolumn{2}{||c|}{Multiplets} & $SU(6)$ representation & $Z_6$ quantum number\\ \hline
& $\chi$ & ${\bf \bar{6}}$ & $+1$\\
fermion & $\hat{\chi}$ & ${\bf\bar{ 6}}$ & $-2$\\
& ${\psi}$ & ${\bf {15}}$ & $+1$\\\hline
 & {$H$} & $\overline{\bf 6} $& $ -2$\\
scalar &${{\hat{H}}}$ &$\bar {\bf 6}$ & { $+1$}\\
&$\Delta$ & ${\bf \overline{15}}$ & $+2$ \\
&$\Sigma$ & {\bf 35} & {0}\\
\hline\hline
\end{tabular}
 \end{center}
 \caption{Quantum numbers of the various fermion and scalar fields under 
the discrete $Z_6$ symmetry in the model of Majorana neutrinos.} 
 \label{tab:II}
 \end{table}
 With this choice of charge assignments the interaction in 
Eq.~(\ref{decouple}) that gives GUT-scale masses to the extra fermions 
$(\hat{L}$, $\hat{D}^c)$ in $\hat{\chi}$ and $(\hat{L}^c$, $\hat{D})$ in 
$\psi$ is allowed by the discrete $Z_6$ symmetry.
The Yukawa couplings that generate masses for the SM quarks 
and charged leptons, Eqs.~(\ref{SMfermionmass}) and 
(\ref{SMfermionmassCorrections}), are also allowed. Turning our 
attention to the neutrino sector, the renormalizable Dirac mass term for 
the neutrinos, Eq.~(\ref{Dmass}), and the nonrenormalizable Majorana 
mass term for the right-handed neutrinos, Eq.~(\ref{Mmass}), are both 
consistent with the discrete symmetry. In the absence of other mass 
terms involving $\nu$ and $\nu^c$, these interactions lead to the 
desired pattern of Majorana neutrino masses. The singlet neutrinos $N$ 
in $\hat{\chi}$ obtain large Majorana masses of order the right-handed 
scale through the operator
 \begin{equation}
-{\cal L}_{\rm RHN} \ = \ \frac{\lambda_{N, ij}}{M_{\rm Pl}} \hat{H}^\dagger \hat{\chi}_i \hat{H} ^\dagger \hat{\chi}_j \; .
\label{Nmass}
 \end{equation}
 The discrete symmetry forbids a renormalizable Dirac mass term between 
the SM neutrinos $\nu$ and the singlet neutrinos $N$. Any allowed Dirac 
mass terms between $\nu^c$ and $N$ are highly Planck suppressed and much 
smaller than their Majorana masses. It follows that the effects of $N$ 
on the neutrino masses are small and can be neglected. Then, the Dirac 
mass term in Eq.~(\ref{Dmass}) and the Majorana mass term in 
Eq.~(\ref{Mmass}) give the dominant contributions to the neutrino 
masses, leading to Majorana neutrino masses of parametrically the right 
size that exhibit the pattern discussed in Section II.

\color{black}
\begin{table}[t!]
\centering
\begin{tabular}{||c|c|c|c|c|c|c|c|c|c|c||}
\hline \hline
Fit  & $y_1$ & $y_2$ & $|y_{11}|$ & $y_{22}$ & $y_{12}$ & $y_{13}$ & $y_{23}$  & $\vartheta$ & $M_0$ (eV)   \\ \hline \hline
Fit 1 (IH)  & 4.152 & 5.100 & 0.9937 & 0.8351 & $-0.0640$ & 0.0537 & 0.0877 & $131.5^\circ$ &$8.485 \times 10^{-4}$ \\ \hline
Fit 2 (IH)  & 4.459 & 4.868 & 0.9773 & 0.8608 & $-0.0624$ & 0.0458 & 0.0745 & $148.0^\circ$  &$1.000 \times 10^{-3}$ \\ \hline
Fit 3 (NH) & 0.5116 & 0.4549 & 0.1330 & -0.7430 & $-0.0375$ & 0.0990  & 0.0263 & $241.3^\circ$&$1.127 \times 10^{-2} $ \\ \hline
Fit 4 (NH) & 0.4983 & 0.4614 & 0.1211 & -0.6934 & $-0.0430$  & 0.0980 & 0.0425 & $245.4^\circ$  &$1.204 \times 10^{-2} $ \\
     \hline \hline
\end{tabular}
\vspace{0.1cm}
 \caption{Values of the parameters chosen for four different benchmark 
models that fit the neutrino oscillation data in the case of Majorana 
neutrinos.}
\vspace{1cm}
 \label{nuTab}

\begin{tabular}{||c|c|c|c|c|c||}
\hline \hline
 {{Oscillation}}   &   {{$3\sigma$ allowed range}} &  \multicolumn{4}{c|}{{Model prediction}}\\
 \cline{3-6}
 {parameters} & {from {\tt NuFit4.1}}~\cite{Esteban:2018azc} & Fit 1 (IH) &  Fit 2 (IH) &  Fit 3 (NH) & Fit 4 (NH) \\ \hline \hline
 $\Delta m_{21}^2 (10^{-5}~{\rm eV}^2$)  &   6.79 - 8.01   &  7.40 & 7.39  & 7.24 & 7.50 \\ \hline
 $\Delta m_{23}^2 (10^{-3}~{\rm eV}^2) $(IH) &   2.416 - 2.603  & 2.509 &  2.504 & - & -\\ 
 $\Delta m_{31}^2 (10^{-3}~{\rm eV}^2) $(NH) &   2.432 - 2.618   & - & -  & 2.532 & 2.500 \\ \hline
  $\sin^2{\theta_{12}}$   &   0.275 - 0.350 &   0.309 &  0.310 & 0.303 & 0.300 \\ \hline
 $\sin^2{\theta_{23}}$  (IH) &   0.430 - 0.612  &   0.590 &  0.544 & - & - \\
 $\sin^2{\theta_{23}}$ (NH)  &   0.427 - 0.609  &  -  &  - & 0.516 & 0.527 \\ \hline 
  $\sin^2{\theta_{13}} $  (IH) &   0.02066 - 0.02461  &   0.02258 &  0.02241 & - & -\\
  $\sin^2{\theta_{13}}  $(NH)  &   0.02046 - 0.02440  &  -  &-   & 0.02232 & 0.02231 \\ \hline
  $\delta_{C\!P}~(^\circ)$ (IH) & 205 - 354 & 296.3 & 286.4 & - & - \\   
  $\delta_{C\!P}~(^\circ)$ (NH) & 141 - 370 & - & - & 282.3 & 277.2 \\ \hline
\end{tabular}
 \caption{Predictions of the benchmark models for the neutrino 
oscillation parameters in the case of Majorana neutrinos, compared to 
the 3$\sigma$ allowed range from a recent global fit.}
 \label{nuTab2}
\end{table}
\begin{figure}[t]
$$\includegraphics[scale=0.3]{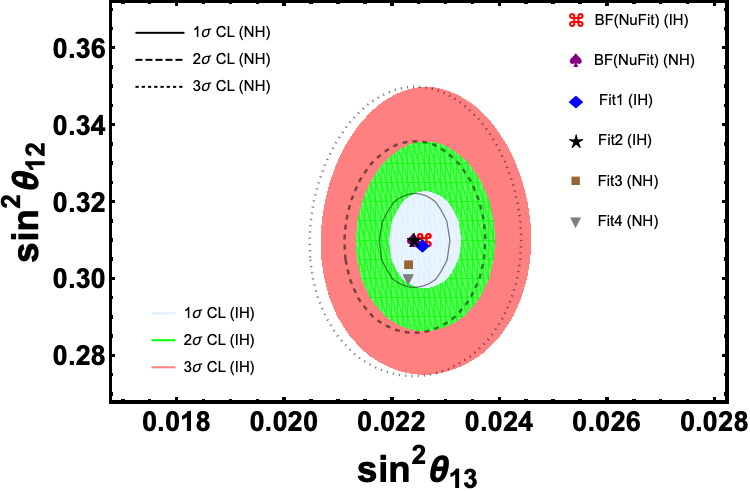} \hspace{3mm}
\includegraphics[scale=0.3]{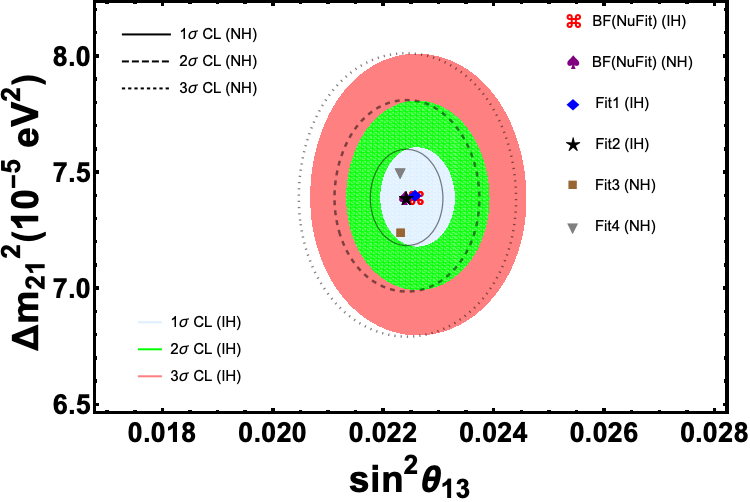} $$
$$ \includegraphics[scale=0.3]{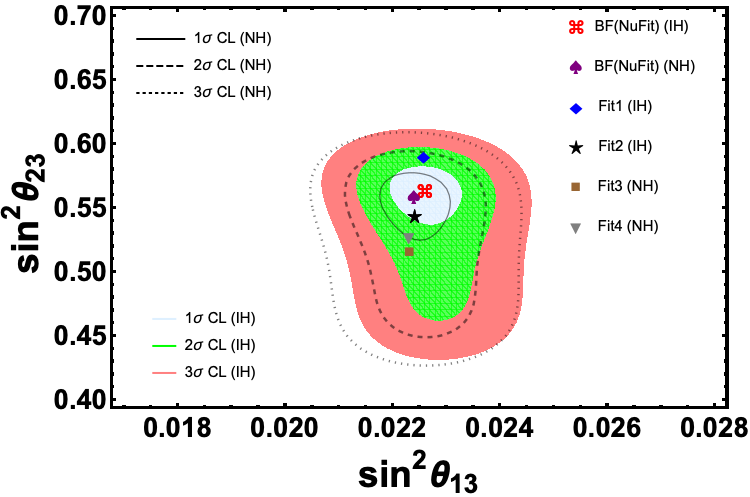} \hspace{3mm}
\includegraphics[scale=0.3]{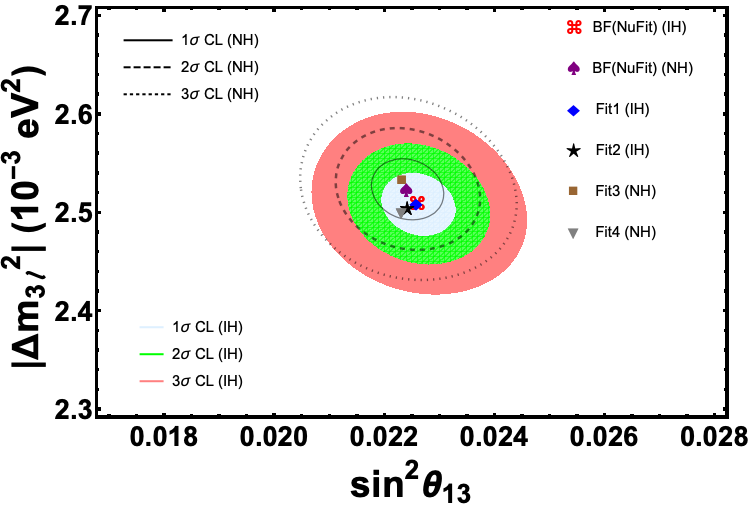}  $$
 \caption{Global oscillation analysis obtained from {\tt NuFit4.1} for 
both the normal hierarchy (NH) and inverted hierarchy (IH) compared to 
our benchmark models for the Majorana case (Fit1, Fit2, Fit3, Fit4). The 
gray, green, and pink-colored contours represent the NuFit $1\sigma, 
2\sigma, \text{and} \, 3\sigma $ CL contours respectively in the NH 
case, whereas the solid, dashed, and dotted lines correspond to the 
$1\sigma, 2\sigma, \text{and} \, 3\sigma $ CL contours respectively for 
IH. The red and purple markers in each case correspond to the NuFit 
best-fit values for the IH and NH respectively, while the blue, black, 
brown, and gray markers are the predictions of the benchmark models 
corresponding to Fit 1, 2, 3, and 4 respectively, as given in 
Table~\ref{nuTab2}. In the bottom right panel, $|\Delta m_{3l}^2|$ 
refers to the atmospheric mass-squared splitting, with $l=1~(2)$ for NH 
(IH).}
 \label{fig:nufit}
\end{figure}
\subsection{Fits to the data }\label{subsec:fitMaj}
\vspace{-4mm}

In this subsection, we obtain fits to the neutrino masses and mixings 
for the case of Majorana neutrinos. The skew-symmetric Dirac mass matrix 
$M_D$ and symmetric Majorana mass matrix $M_{\nu^c}$ are parameterized 
as
 \begin{equation} M_D 
 \ = \ \begin{pmatrix}
    	0 & m_1 & m_2 \\
    	-m_1 & 0 & m_3 \\
    	-m_2 & -m_3 & 0 
	\end{pmatrix}, \qquad 
M_{\nu^c}  \ = \ \begin{pmatrix}
   				 M_{11} & M_{12}  & M_{13} \\
   				M_{12}  & M_{22} & M_{23} \\
    			M_{13} &  M_{23} & M_{33} 
			\end{pmatrix} \, .
\label{eq:MajMat}
\end{equation}
 In the limit that $M_D \ll M_\nu^c$, we can write the following seesaw 
relation for the light neutrino masses,
 \begin{align}
M_\nu  \ & \simeq \  -M_D M_{\nu^c}^{-1} M_D^T \nonumber \\
 \ & = \ - M_0 \begin{pmatrix}
    	0 & y_1 & y_2 \\
    	-y_1 & 0 & 1 \\
    	-y_2 & -1 & 0 
	\end{pmatrix} 
	\begin{pmatrix}
   				 |y_{11}|e^{i\vartheta} & y_{12}  & y_{13} \\
   				y_{12}  & y_{22} & y_{23} \\
    			y_{13} &  y_{23} & 1 
	\end{pmatrix}^{-1}
	\begin{pmatrix}
    	0 & -y_1 & -y_2 \\
    	y_1 & 0 & -1 \\
    	y_2 & 1 & 0 
	\end{pmatrix} \, ,
\label{eq:seesaw2}
\end{align}
 where we choose to parametrize the mass matrix in terms of $y_i \equiv 
{m_i}/{m_3}$, $y_{ij} \equiv {M_{ij}}/{M_{33}}$, and $M_0 \equiv 
{m_3^2}/{M_{33}}$. The overall mass scale $M_0$ is required to be tiny, 
of order $10^{-11}$ GeV, to obtain the observed values of neutrino 
masses. We perform a numerical scan of the input parameters, as shown in 
Eq.~\eqref{eq:seesaw2}, to obtain predictions for the entire PMNS 
matrix. It is beyond the scope of this work to scan over the full 
parameter space; instead, we perform a constrained minimization in which 
the five neutrino observables ($\sin^2 \theta_{12}, \sin^2 \theta_{13}, 
\sin^2 \theta_{23}, \Delta m_{21}^2$, and $|\Delta m_{3l}^2|$ with $l=1$ 
in the case of normal hierarchy and $l=2$ for inverted) are restricted 
to lie within $2 \sigma$ of their experimentally measured values. The 
parameter $M_{11}$ has been chosen to be complex in order to induce a 
$C\!P$ violating phase in the PMNS matrix, but the other parameters have 
been taken to be real. We emphasize that the lightest neutrino is 
exactly massless due to the skew-symmetric nature of the Dirac mass 
matrix $M_D$.

The input parameters shown in Table \ref{nuTab} provide excellent fits 
to the oscillation data, as can be seen in Table \ref{nuTab2}. For 
each of the benchmark points the $C\!P$ phase in the PMNS matrix is 
large, showing that there is no restriction on its value. Fits 1 and 2 
correspond to an inverted hierarchy, whereas Fits 3 and 4 represent a 
normal hierarchy. The benchmark points (Fit 1, Fit 2, Fit 3 and Fit 4) 
are also displayed in Fig.~\ref{fig:nufit} as Fit1 (IH), Fit2 (IH), Fit3 
(NH), and Fit4 (NH) as blue, black, brown, and gray markers respectively 
in various two-dimensional projections of the global-fit 
results~\cite{Esteban:2018azc}.


\subsection{Neutrinoless double beta Decay}
 In the standard framework with only light neutrinos contributing to 
$0\nu\beta\beta$, the amplitude for the $0\nu\beta\beta$ rate is 
proportional to the $ee-$element of the neutrino mass matrix, given by
 \begin{equation}
    m_{ee}  \ = \ |m_1 c_{12}^2 c_{13}^2 + e^{ i \alpha} m_2 s_{12}^2 c_{13}^2 + e^{ i \beta} m_3 s_{13}^2| \;.
    \label{eq:doublebeta}
 \end{equation}
 Here $m_1$, $m_2$, and $m_3$ are the masses of the three light 
neutrinos, while $s_{ij}^2 \equiv \sin^2{\theta_{ij}}$, $c_{ij}^2 \equiv 
\cos^2{\theta_{ij}}$ (for $ij = 12, 13, 23$), and ($\alpha$, $\beta$) 
are the two unknown Majorana phases.

We can apply Eq.~(\ref{eq:doublebeta}) to our framework to determine its 
implications for $0\nu\beta\beta$. Since the determinant of $M_D$ 
vanishes owing to its skew-symmetric structure, the lightest neutrino is 
exactly massless. For a given mass ordering (normal or inverted), the 
masses of the heavier two neutrinos can then be determined from the 
observed mass splittings. The expression for the effective Majorana mass 
given in Eq.~(\ref{eq:doublebeta}) then reduces to one of the following 
equations, depending on whether the hierarchy is normal or inverted:
 \begin{eqnarray}
 m_{ee}^{\rm NH} & \ = \ & \left|\sqrt{\Delta m_{21}^2} s_{12}^2 c_{13}^2 + \sqrt{\Delta m_{31}^2} s_{13}^2 e^{i (\beta-\alpha)} \right| \, , \label{eq:meeNH}\\
 m_{ee}^{\rm IH} & \ = \ & \left|\sqrt{|\Delta m_{32}^2|-\Delta m_{21}^2}  \, c_{12}^2 c_{13}^2 + \sqrt{|\Delta m_{32}^2|} \, s_{12}^2 c_{13}^2  e^{i \alpha} \right| \, .
  \label{eq:meeIH}
 \end{eqnarray}
 Note that only one Majorana phase (or one specific linear combination 
of phases) is relevant, due to the smallest mass eigenvalue being zero.

\begin{figure}
    \centering
    \includegraphics[scale=0.33]{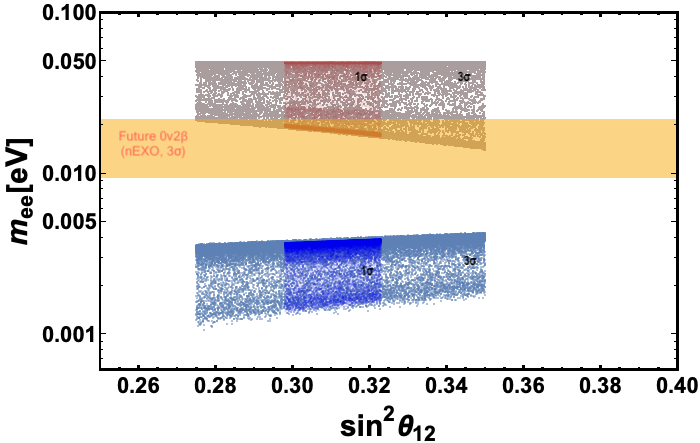}
    \includegraphics[scale=0.33]{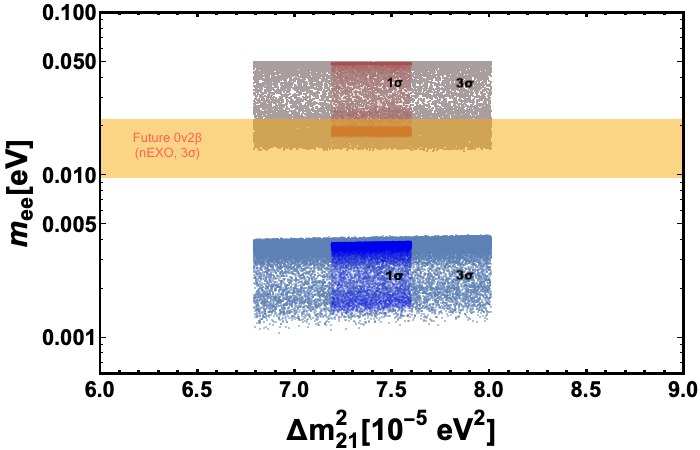} \\ \vspace{4mm}
    \includegraphics[scale=0.33]{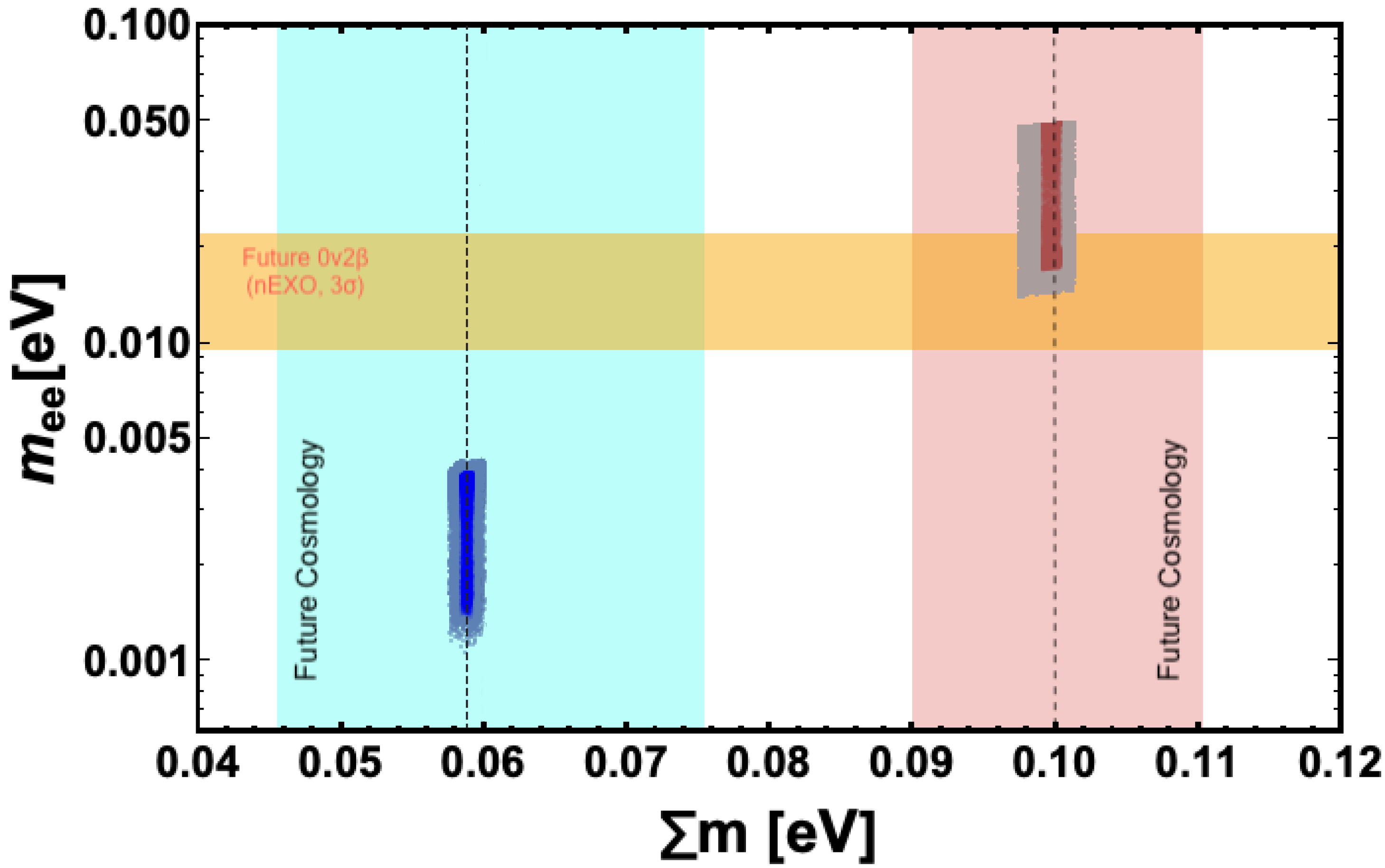}
    \caption{Model predictions for the effective Majorana mass $m_{ee}$ 
as a function of $\sin^2\theta_{12}$ (left), $\Delta m_{21}^2$ (right), 
and $\sum m_i$ (bottom). The blue (red) points correspond to NH (IH) and 
the dark (light) color corresponds to the $1\sigma$ ($3\sigma$) CL for 
the oscillation observables. The horizontal orange band shows the 
sensitivity of the future $0\nu\beta\beta$ experiment nEXO at $3\sigma$ 
CL. The vertical blue (red) band shows the forecast $1\sigma$ limits on 
$\sum m_i$ from CMB-S4 in the case of NH (IH), whereas the vertical 
dotted lines show the corresponding central values.}
    \label{fig:obvv}
\end{figure}

To illustrate the range of possibilities for $0\nu\beta\beta$ in this 
class of models, in Fig.~\ref{fig:obvv} we plot the effective Majorana 
mass as a function of $\sin^2\theta_{12}$, $\Delta m_{21}^2$ and the sum 
of light neutrino masses $\sum m_i$. We restrict to points that lie 
within $1\sigma$ and $3\sigma$ of the allowed oscillation parameter 
range. Each data point in Fig.~\ref{fig:obvv} represents a valid fit 
that has been obtained by scanning over the input parameters shown in 
Eq.~(\ref{eq:seesaw2}). For the purposes of this scan, we have taken all 
the elements of the $M_{\nu^c}$ matrix to be complex. Here the blue 
(red) points correspond to the case of normal (inverted) hierarchy. The 
Majorana phases, as well as the other observables in 
Eqs.~(\ref{eq:meeNH}) and (\ref{eq:meeIH}), have been obtained as 
predictions of the points in the scan. First, the PMNS matrix is 
identified with the matrix diagonalizing $M_\nu^\dagger M_\nu$, where 
$M_\nu$ is given in Eq.~\eqref{eq:seesaw2}. Then, taking $U^T M_\nu 
U=D_\nu$ gives the diagonalized mass matrix with the appropriate 
Majorana phases.

We can use Eqs.~(\ref{eq:meeNH}) and (\ref{eq:meeIH}) to obtain upper 
and lower limits on the rate of $0\nu\beta\beta$ in this 
class of models. In the case of a normal hierarchy, the two terms in 
Eq.~(\ref{eq:meeNH}) add constructively for $0 \leq (\beta-\alpha) \leq 
\pi /2$, while partial cancellation occurs for $\pi/2 \leq 
(\beta-\alpha) \leq \pi$. The most effective cancellation (addition) 
happens when $\beta-\alpha = \pi \, (0)$. This allows us to 
calculate the minimum and maximum values of the effective Majorana mass, 
which is parameterized as
 \begin{equation}
    m_{ee}^{\rm{MIN, MAX}}~({\rm NH}) \ = \ \left|\sqrt{\Delta m_{21}^2} s_{12}^2 c_{13}^2 \mp \sqrt{\Delta m_{31}^2} s_{13}^2 \right| \, .
 \end{equation}
 Allowing the fit values from ${\tt NuFit4.1}$ to vary over the 
$3\sigma$ range, the minimum effective Majorana mass is obtained as 
$m_{ee}^{\rm MIN} = 9.7 \times 10^{-4}$ eV, whereas the maximum 
effective Majorana mass is $m_{ee}^{\rm{MAX}} = 4.3 \times 
10^{-3}~\rm{eV}$. One can make similar arguments in the case of an
inverted hierarchy, for which the most effective cancellation (enhancement) happens when $\alpha
= \pi~(0)$ in Eq.~(\ref{eq:meeIH}). This leads to
 \begin{equation}
    m_{ee}^{\rm{MIN, MAX}}~({\rm IH}) \ = \ \left|\sqrt{|\Delta m_{32}^2|-\Delta m_{21}^2}  \, c_{12}^2 c_{13}^2 \mp \sqrt{|\Delta m_{32}^2|} \, s_{12}^2 c_{13}^2   \right| \, ,
 \end{equation}
 This allows us to determine the minimum and maximum values of the 
effective Majorana mass in the case of an inverted mass hierarchy as 
$m_{ee}^{\rm MIN} = 1.39 \times 10^{-2}$ eV and $m_{ee}^{\rm MAX} = 4.95 
\times 10^{-2}$ eV respectively.

Future ton-scale $0\nu\beta\beta$ experiments such as 
LEGEND~\cite{Abgrall:2017syy} and nEXO~\cite{Albert:2017hjq} should be 
able to probe the entire parameter space of this class of models if the 
hierarchy is inverted. For illustration, we show in Fig.~\ref{fig:obvv} 
the future sensitivity from nEXO~\cite{Albert:2017hjq} at $3\sigma$ CL 
(horizontal orange band), where the band takes into account the nuclear 
matrix element uncertainties involved in translating a given lower bound 
on the half-life into an upper bound on the effective Majorana mass 
parameter.

Similarly, a future cosmological measurement of the sum of the light 
neutrino masses $\sum m_i$ would allow another test of the model 
predictions. Shown in the bottom panel of Fig.~\ref{fig:obvv} are the 
$1\sigma$ sensitivity from CMB-S4~\cite{Abazajian:2019eic} (vertical 
band) for both the normal hierarchy (blue) and inverted hierarchy (red). 
It is clear from the figure that the model predictions lie well within 
the $1\sigma$ sensitivity of CMB-S4, and so these measurements offer an 
opportunity to test this scenario.

\section{Conclusion} \label{sec:V}

In summary, we have presented a framework for neutrino masses in $SU(6)$ 
GUTs that predicts a specific texture for the form of the leading 
contribution to the Dirac mass term. In this scenario, neutrinos can be 
either Dirac or Majorana particles. A concrete prediction in the Dirac 
case is that the mass hierarchy is inverted. In the Majorana case, on 
the other hand, both normal and inverted hierarchies are allowed. In 
both the Dirac and Majorana cases, the model makes cosmologically 
testable predictions regarding the sum of neutrino masses. Furthermore, 
in the case of Majorana neutrinos, this framework predicts lower and 
upper bounds on the rate of $0\nu\beta\beta$ for both the normal and the 
inverted hierarchies. In the case of an inverted hierarchy, this 
prediction can be tested in future ton-scale $0\nu\beta\beta$ 
experiments.

\noindent
 {\bf Note Added:} While this work was in progress we received 
Ref.~\cite{Li:2019qxy}, which considers Majorana neutrino masses in the 
context of an intermediate scale $SU(3) \times SU(3) \times U(1)$ model 
embedded in an $SU(6)$ GUT. Although based on the inverse seesaw 
framework, the resulting pattern of neutrino masses shares some of the 
features of our Majorana construction, including the skew-symmetric form 
of the Dirac mass term and a massless neutrino.

 \section*{Acknowledgments} 
 The work of ZC and RNM is supported in part by the National Science 
Foundation under Grant No. PHY-1914731.  The work of BD is supported in 
part by the U.S. Department of Energy under Grant No. DE-SC0017987 and 
in part by the MCSS funds.  The work of AT is supported in part by the US 
Department of Energy Grant Number DE-SC0016013. ZC is also supported in 
part by the US-Israeli BSF grant 2018236. The work of BD and AT was also supported 
by the Neutrino Theory Network Program under Grant 
No.~DE-AC02-07CH11359. BD acknowledges the High Energy Theory 
group at Oklahoma State University for warm hospitality, where part of this work was 
performed.

\bibliographystyle{utcaps_mod}
\bibliography{references}

\providecommand{\href}[2]{#2}\begingroup\raggedright\begin{thebibliography}{10}

\bibitem{Tanabashi:2018oca}
{\normalfont \bfseries Particle Data Group}, M.~Tanabashi {\em et al.}, ``{\em
  {Review of Particle Physics}},''
\href{http://dx.doi.org/10.1103/PhysRevD.98.030001}{Phys. Rev. {\normalfont
  \bfseries D98} (2018) no.~3, 030001}.

\bibitem{Qian:2015waa}
X.~Qian and P.~Vogel, ``{\em {Neutrino Mass Hierarchy}},''
  \href{http://dx.doi.org/10.1016/j.ppnp.2015.05.002}{Prog. Part. Nucl. Phys.
  {\normalfont \bfseries 83} (2015)  1--30},
  \href{http://arxiv.org/abs/1505.01891}{{\normalfont \ttfamily
  arXiv:1505.01891}}.

\bibitem{Dolinski:2019nrj}
M.~J. Dolinski, A.~W. Poon, and W.~Rodejohann, ``{\em {Neutrinoless Double-Beta
  Decay: Status and Prospects}},''
  \href{http://dx.doi.org/10.1146/annurev-nucl-101918-023407}{Ann. Rev. Nucl.
  Part. Sci. {\normalfont \bfseries 69} (2019)  219--251},
  \href{http://arxiv.org/abs/1902.04097}{{\normalfont \ttfamily
  arXiv:1902.04097}}.

\bibitem{Pati:1974yy}
J.~C. Pati and A.~Salam, ``{\em {Lepton Number as the Fourth Color}},''
  \href{http://dx.doi.org/10.1103/PhysRevD.10.275}{Phys. Rev. D {\normalfont
  \bfseries 10} (1974)  275--289}. [Erratum: Phys.Rev.D 11, 703--703 (1975)].

\bibitem{Georgi:1974sy}
H.~Georgi and S.~Glashow, ``{\em {Unity of All Elementary Particle Forces}},''
  \href{http://dx.doi.org/10.1103/PhysRevLett.32.438}{Phys. Rev. Lett.
  {\normalfont \bfseries 32} (1974)  438--441}.

\bibitem{Langacker:1980js}
P.~Langacker, ``{\em {Grand Unified Theories and Proton Decay}},''
  \href{http://dx.doi.org/10.1016/0370-1573(81)90059-4}{Phys. Rept.
  {\normalfont \bfseries 72} (1981)  185}.

\bibitem{Buras:1977yy}
A.~Buras, J.~R. Ellis, M.~Gaillard, and D.~V. Nanopoulos, ``{\em {Aspects of
  the Grand Unification of Strong, Weak and Electromagnetic Interactions}},''
  \href{http://dx.doi.org/10.1016/0550-3213(78)90214-6}{Nucl. Phys. B
  {\normalfont \bfseries 135} (1978)  66--92}.

\bibitem{Georgi:1974my}
H.~Georgi, ``{\em {The State of the Art---Gauge Theories}},''
  \href{http://dx.doi.org/10.1063/1.2947450}{AIP Conf. Proc. {\normalfont
  \bfseries 23} (1975)  575--582}.

\bibitem{Fritzsch:1974nn}
H.~Fritzsch and P.~Minkowski, ``{\em {Unified Interactions of Leptons and
  Hadrons}},'' \href{http://dx.doi.org/10.1016/0003-4916(75)90211-0}{Annals
  Phys. {\normalfont \bfseries 93} (1975)  193--266}.

\bibitem{Kim:1981jw}
J.~E. Kim, ``{\em {Reason for SU(6) Grand Unification}},''
  \href{http://dx.doi.org/10.1016/0370-2693(81)91149-7}{Phys. Lett. B
  {\normalfont \bfseries 107} (1981)  69--72}.

\bibitem{Fukugita:1981gn}
M.~Fukugita, T.~Yanagida, and M.~Yoshimura, ``{\em {N anti-N Oscillation
  without Left-Right Symmetry}},''
  \href{http://dx.doi.org/10.1016/0370-2693(82)91092-9}{Phys. Lett. B
  {\normalfont \bfseries 109} (1982)  369--372}.

\bibitem{Abell:2009aa}
{\normalfont \bfseries LSST}, P.~A. Abell {\em et al.}, ``{\em {LSST Science
  Book, Version 2.0}},'' \href{http://arxiv.org/abs/0912.0201}{{\normalfont
  \ttfamily arXiv:0912.0201}}.

\bibitem{Amendola:2016saw}
L.~Amendola {\em et al.}, ``{\em {Cosmology and fundamental physics with the
  Euclid satellite}},''
  \href{http://dx.doi.org/10.1007/s41114-017-0010-3}{Living Rev. Rel.
  {\normalfont \bfseries 21} (2018) no.~1, 2},
  \href{http://arxiv.org/abs/1606.00180}{{\normalfont \ttfamily
  arXiv:1606.00180}}.

\bibitem{Aghamousa:2016zmz}
{\normalfont \bfseries DESI}, A.~Aghamousa {\em et al.}, ``{\em {The DESI
  Experiment Part I: Science,Targeting, and Survey Design}},''
  \href{http://arxiv.org/abs/1611.00036}{{\normalfont \ttfamily
  arXiv:1611.00036}}.

\bibitem{Ade:2018sbj}
{\normalfont \bfseries Simons Observatory}, P.~Ade {\em et al.}, ``{\em {The
  Simons Observatory: Science goals and forecasts}},''
  \href{http://dx.doi.org/10.1088/1475-7516/2019/02/056}{JCAP {\normalfont
  \bfseries 02} (2019)  056},
  \href{http://arxiv.org/abs/1808.07445}{{\normalfont \ttfamily
  arXiv:1808.07445}}.

\bibitem{Abazajian:2019eic}
{\normalfont \bfseries CMB-S4}, K.~Abazajian {\em et al.}, ``{\em {CMB-S4
  Science Case, Reference Design, and Project Plan}},''
  \href{http://arxiv.org/abs/1907.04473}{{\normalfont \ttfamily
  arXiv:1907.04473}}.

\bibitem{Esfahani:2017dmu}
{\normalfont \bfseries Project 8}, A.~Ashtari~Esfahani {\em et al.}, ``{\em
  {Determining the neutrino mass with cyclotron radiation emission
  spectroscopy---Project 8}},''
  \href{http://dx.doi.org/10.1088/1361-6471/aa5b4f}{J. Phys. G {\normalfont
  \bfseries 44} (2017) no.~5, 054004},
  \href{http://arxiv.org/abs/1703.02037}{{\normalfont \ttfamily
  arXiv:1703.02037}}.

\bibitem{Abe:2018uyc}
{\normalfont \bfseries Hyper-Kamiokande}, K.~Abe {\em et al.}, ``{\em
  {Hyper-Kamiokande Design Report}},''
  \href{http://arxiv.org/abs/1805.04163}{{\normalfont \ttfamily
  arXiv:1805.04163}}.

\bibitem{Abi:2018dnh}
{\normalfont \bfseries DUNE}, B.~Abi {\em et al.}, ``{\em {The DUNE Far
  Detector Interim Design Report Volume 1: Physics, Technology and
  Strategies}},'' \href{http://arxiv.org/abs/1807.10334}{{\normalfont \ttfamily
  arXiv:1807.10334}}.

\bibitem{Vissani:1999tu}
F.~Vissani, ``{\em {Signal of neutrinoless double beta decay, neutrino spectrum
  and oscillation scenarios}},''
  \href{http://dx.doi.org/10.1088/1126-6708/1999/06/022}{JHEP {\normalfont
  \bfseries 06} (1999)  022},
  \href{http://arxiv.org/abs/hep-ph/9906525}{{\normalfont \ttfamily
  arXiv:hep-ph/9906525}}.

\bibitem{Bilenky:2001rz}
S.~M. Bilenky, S.~Pascoli, and S.~Petcov, ``{\em {Majorana neutrinos, neutrino
  mass spectrum, CP violation and neutrinoless double beta decay. 1. The Three
  neutrino mixing case}},''
  \href{http://dx.doi.org/10.1103/PhysRevD.64.053010}{Phys. Rev. D {\normalfont
  \bfseries 64} (2001)  053010},
  \href{http://arxiv.org/abs/hep-ph/0102265}{{\normalfont \ttfamily
  arXiv:hep-ph/0102265}}.

\bibitem{Mohapatra:2005wg}
R.~N. Mohapatra {\em et al.}, ``{\em {Theory of neutrinos: A White paper}},''
  \href{http://dx.doi.org/10.1088/0034-4885/70/11/R02}{Rept. Prog. Phys.
  {\normalfont \bfseries 70} (2007)  1757--1867},
  \href{http://arxiv.org/abs/hep-ph/0510213}{{\normalfont \ttfamily
  arXiv:hep-ph/0510213}}.

\bibitem{Sen:1984aq}
A.~Sen, ``{\em {Sliding Singlet Mechanism in N=1 Supergravity GUT}},''
  \href{http://dx.doi.org/10.1016/0370-2693(84)91612-5}{Phys. Lett. B
  {\normalfont \bfseries 148} (1984)  65--68}.

\bibitem{Barr:1997pt}
S.~M. Barr, ``{\em {The Sliding - singlet mechanism revived}},''
  \href{http://dx.doi.org/10.1103/PhysRevD.57.190}{Phys. Rev. D {\normalfont
  \bfseries 57} (1998)  190--194},
  \href{http://arxiv.org/abs/hep-ph/9705266}{{\normalfont \ttfamily
  arXiv:hep-ph/9705266}}.

\bibitem{Berezhiani:1989bd}
Z.~Berezhiani and G.~Dvali, ``{\em {Possible solution of the hierarchy problem
  in supersymmetrical grand unification theories}},'' Bull. Lebedev Phys. Inst.
  {\normalfont \bfseries 5} (1989)  55--59.

\bibitem{Barbieri:1993wz}
R.~Barbieri, G.~Dvali, and M.~Moretti, ``{\em {Back to the doublet - triplet
  splitting problem}},''
  \href{http://dx.doi.org/10.1016/0370-2693(93)90501-8}{Phys. Lett. B
  {\normalfont \bfseries 312} (1993)  137--142}.

\bibitem{Dvali:1993yf}
G.~Dvali, ``{\em {Why is the Higgs doublet light?}},''
  \href{http://dx.doi.org/10.1016/0370-2693(94)00075-1}{Phys. Lett. B
  {\normalfont \bfseries 324} (1994)  59--65}.

\bibitem{Chacko:1998zz}
Z.~Chacko and R.~N. Mohapatra, ``{\em {Doublet triplet splitting in
  supersymmetric SU(6) by missing VEV mechanism}},''
  \href{http://dx.doi.org/10.1016/S0370-2693(98)01263-5}{Phys. Lett. B
  {\normalfont \bfseries 442} (1998)  199--202},
  \href{http://arxiv.org/abs/hep-ph/9809345}{{\normalfont \ttfamily
  arXiv:hep-ph/9809345}}.

\bibitem{Barbieri:1994kw}
R.~Barbieri, G.~Dvali, A.~Strumia, Z.~Berezhiani, and L.~J. Hall, ``{\em
  {Flavor in supersymmetric grand unification: A Democratic approach}},''
  \href{http://dx.doi.org/10.1016/0550-3213(94)90593-2}{Nucl. Phys. B
  {\normalfont \bfseries 432} (1994)  49--67},
  \href{http://arxiv.org/abs/hep-ph/9405428}{{\normalfont \ttfamily
  arXiv:hep-ph/9405428}}.

\bibitem{Berezhiani:1995dt}
Z.~Berezhiani, ``{\em {SUSY SU(6) GIFT for doublet-triplet splitting and
  fermion masses}},''
  \href{http://dx.doi.org/10.1016/0370-2693(95)00705-P}{Phys. Lett. B
  {\normalfont \bfseries 355} (1995)  481--491},
  \href{http://arxiv.org/abs/hep-ph/9503366}{{\normalfont \ttfamily
  arXiv:hep-ph/9503366}}.

\bibitem{Minkowski:1977sc}
P.~Minkowski, ``{\em {$\mu \to e\gamma$ at a Rate of One Out of $10^{9}$ Muon
  Decays?}},'' \href{http://dx.doi.org/10.1016/0370-2693(77)90435-X}{Phys.
  Lett. B {\normalfont \bfseries 67} (1977)  421--428}.

\bibitem{Mohapatra:1979ia}
R.~N. Mohapatra and G.~Senjanovic, ``{\em {Neutrino Mass and Spontaneous Parity
  Nonconservation}},''
  \href{http://dx.doi.org/10.1103/PhysRevLett.44.912}{Phys. Rev. Lett.
  {\normalfont \bfseries 44} (1980)  912}.

\bibitem{Yanagida:1979as}
T.~Yanagida, ``{\em {Horizontal gauge symmetry and masses of neutrinos}},''
  Conf. Proc. C {\normalfont \bfseries 7902131} (1979)  95--99.

\bibitem{GellMann:1980vs}
M.~Gell-Mann, P.~Ramond, and R.~Slansky, ``{\em {Complex Spinors and Unified
  Theories}},'' Conf. Proc. C {\normalfont \bfseries 790927} (1979)  315--321,
  \href{http://arxiv.org/abs/1306.4669}{{\normalfont \ttfamily
  arXiv:1306.4669}}.

\bibitem{Esteban:2018azc}
I.~Esteban, M.~Gonzalez-Garcia, A.~Hernandez-Cabezudo, M.~Maltoni, and
  T.~Schwetz, ``{\em {Global analysis of three-flavour neutrino oscillations:
  synergies and tensions in the determination of $\theta_{23}$, $\delta_{CP}$,
  and the mass ordering}},''
  \href{http://dx.doi.org/10.1007/JHEP01(2019)106}{JHEP {\normalfont \bfseries
  01} (2019)  106}, \href{http://arxiv.org/abs/1811.05487}{{\normalfont
  \ttfamily arXiv:1811.05487}}. \url{http://www.nu-fit.org/}.

\bibitem{Abe:2019vii}
{\normalfont \bfseries T2K}, K.~Abe {\em et al.}, ``{\em {Constraint on the
  matter--antimatter symmetry-violating phase in neutrino oscillations}},''
  \href{http://dx.doi.org/10.1038/s41586-020-2177-0}{Nature {\normalfont
  \bfseries 580} (2020) no.~7803, 339--344},
  \href{http://arxiv.org/abs/1910.03887}{{\normalfont \ttfamily
  arXiv:1910.03887}}.

\bibitem{Abgrall:2017syy}
{\normalfont \bfseries LEGEND}, N.~Abgrall {\em et al.}, ``{\em {The Large
  Enriched Germanium Experiment for Neutrinoless Double Beta Decay
  (LEGEND)}},'' \href{http://dx.doi.org/10.1063/1.5007652}{AIP Conf. Proc.
  {\normalfont \bfseries 1894} (2017) no.~1, 020027},
  \href{http://arxiv.org/abs/1709.01980}{{\normalfont \ttfamily
  arXiv:1709.01980}}.

\bibitem{Albert:2017hjq}
{\normalfont \bfseries nEXO}, J.~Albert {\em et al.}, ``{\em {Sensitivity and
  Discovery Potential of nEXO to Neutrinoless Double Beta Decay}},''
  \href{http://dx.doi.org/10.1103/PhysRevC.97.065503}{Phys. Rev. C {\normalfont
  \bfseries 97} (2018) no.~6, 065503},
  \href{http://arxiv.org/abs/1710.05075}{{\normalfont \ttfamily
  arXiv:1710.05075}}.

\bibitem{Li:2019qxy}
T.~Li, J.~Pei, F.~Xu, and W.~Zhang, ``{\em {The $SU(3)_C\times SU(3)_L\times
  U(1)_X$ Model from $SU(6)$}},''
  \href{http://arxiv.org/abs/1911.09551}{{\normalfont \ttfamily
  arXiv:1911.09551}}.

\end{thebibliography}\endgroup


\providecommand{\href}[2]{#2}\begingroup\raggedright\endgroup
\end{document}